\documentclass[11pt]{article}  
\usepackage[margin=1in]{geometry}
\usepackage{multicol,multirow}
\usepackage[utf8]{inputenc}
\usepackage[T1]{fontenc}
\usepackage{lmodern}
\usepackage{booktabs, multirow}
\usepackage{changepage,threeparttable}
\usepackage{amssymb,amsmath,amsthm,amsfonts}
\usepackage{textgreek}
\usepackage{mathtools}
\usepackage{empheq} % loads amsmath, mathtools
\usepackage{enumitem}
\usepackage[numbers,comma,sort&compress]{natbib}
\usepackage{authblk}
\usepackage{graphicx, caption, subcaption}
\usepackage{svg}
\usepackage{float}
\usepackage[ruled,vlined,linesnumbered]{algorithm2e}
\usepackage{algpseudocode}
\usepackage{physics}
\usepackage{footnote}
\usepackage{xcolor}
\usepackage{mathrsfs}
\usepackage{bbm}
\usepackage{bm}
\usepackage{pythonhighlight}
\usepackage{adjustbox}
\usepackage{colortbl}

\usepackage{array}
\newcolumntype{P}[1]{>{\centering\arraybackslash}p{#1}}

\usepackage[colorlinks,linkcolor=cyan,citecolor=cyan,urlcolor=cyan]{hyperref}

\usepackage{siunitx}
\theoremstyle{plain}

\theoremstyle{definition}

\newtheorem{example}{Example}

\usepackage{thm-restate}

\newcommand{\eqn}[1]{\hyperref[eqn:#1]{(\ref*{eqn:#1})}}
\newcommand{\rem}[1]{\hyperref[rem:#1]{Remark~\ref*{rem:#1}}}
\newcommand{\thm}[1]{\hyperref[thm:#1]{Theorem~\ref*{thm:#1}}}
\newcommand{\cor}[1]{\hyperref[cor:#1]{Corollary~\ref*{cor:#1}}}
\newcommand{\defn}[1]{\hyperref[defn:#1]{Definition~\ref*{defn:#1}}}
\newcommand{\assump}[1]{\hyperref[assump:#1]{Assumption~\ref*{assump:#1}}}
\newcommand{\lem}[1]{\hyperref[lem:#1]{Lemma~\ref*{lem:#1}}}
\newcommand{\prop}[1]{\hyperref[prop:#1]{Proposition~\ref*{prop:#1}}}
\newcommand{\fig}[1]{\hyperref[fig:#1]{Figure~\ref*{fig:#1}}}
\newcommand{\tab}[1]{\hyperref[tab:#1]{Table~\ref*{tab:#1}}}
\newcommand{\algo}[1]{\hyperref[algo:#1]{Algorithm~\ref*{algo:#1}}}
\renewcommand{\sec}[1]{\hyperref[sec:#1]{Section~\ref*{sec:#1}}}
\newcommand{\append}[1]{\hyperref[append:#1]{Appendix~\ref*{append:#1}}}
\newcommand{\fac}[1]{\hyperref[fac:#1]{Fact~\ref*{fac:#1}}}
\newcommand{\lin}[1]{\hyperref[lin:#1]{Line~\ref*{lin:#1}}}
\newcommand{\fnote}[1]{\hyperref[fnote:#1]{Footnote~\ref*{fnote:#1}}}

\def\>{\rangle}
\def\<{\langle}
\def\trans{^{\top}}

\newcommand{\R}{\mathbb{R}}

\renewcommand{\i}{\mathrm{i}}

\renewcommand\bra[1]{{\langle{#1}|}}
\makeatletter
\renewcommand\ket[1]{%
  \@ifnextchar\bra{\k@t{#1}\!}{\k@t{#1}}%
}
\newcommand\k@t[1]{{|{#1}\rangle}}
\makeatother

\numberwithin{equation}{section}

\newcommand{\qhdopt}{\texttt{QHDOPT}}
\newcommand{\simuq}{\texttt{SimuQ}}

\begin{document}

\title{QHDOPT: A Software for Nonlinear Optimization with Quantum Hamiltonian Descent}
\author[1] {Samuel Kushnir}
\author[2,3,4,$\dagger$,\footnote{Samuel Kushnir and Jiaqi Leng contributed equally to this work. Most of the work was completed at the University of Maryland.}] {Jiaqi Leng}
\author[1,3] {Yuxiang Peng}
\author[5,6] {Lei Fan}
\author[1,3] {Xiaodi Wu}
\affil[1]{Department of Computer Science, University of Maryland}
\affil[2]{Department of Mathematics, University of Maryland}
\affil[3]{Joint Center for Quantum Information and Computer Science, University of Maryland}
\affil[4]{Department of Mathematics and Simons Institute for the Theory of Computing, University of California, Berkeley}
\affil[5]{Department of Engineering Technology, University of Houston}
\affil[6]{Department of Electrical and Computer Engineering, University of Houston}
\affil[$\dagger$]{\href{mailto:jiaqil@terpmail.umd.edu}{jiaqil@terpmail.umd.edu}}
\date{}

\maketitle

\begin{abstract}
We develop an open-source, end-to-end software (named QHDOPT), which can solve nonlinear optimization problems using the quantum Hamiltonian descent (QHD) algorithm. 
QHDOPT offers an accessible interface and automatically maps tasks to various supported quantum backends (i.e., quantum hardware machines). 
These features enable users, even those without prior knowledge or experience in quantum computing, to utilize the power of existing quantum devices for nonlinear and nonconvex optimization tasks.
In its intermediate compilation layer, QHDOPT employs SimuQ, an efficient interface for Hamiltonian-oriented programming, to facilitate multiple algorithmic specifications and ensure compatible cross-hardware deployment.
The detailed documentation of QHDOPT is available at \url{https://github.com/jiaqileng/QHDOPT}.
\end{abstract}

\section{Introduction}
Nonlinear optimization, also known as nonlinear programming, is a branch of mathematical optimization concerned with solving problems in which the objective function, constraints, or both, exhibit nonlinearity. While nonlinear optimization problems are common in various application fields such as engineering, management, economics, and finance, these problems are in general nonconvex with complicated landscape features like multiple local stationary points, valleys, and plateaus. As the number of variables grows, the complexity of the problem could grow rapidly, posing a significant challenge in obtaining globally optimal solutions.

Several open-source and commercial software packages, including Ipopt~\citep{kawajir2006introduction}, Gurobi~\citep{gurobi2021gurobi}, and CPLEX~\citep{bliek1u2014solving}, have been developed to tackle large-scale nonlinear optimization problems. While these optimizers can incorporate powerful heuristics to enhance performance for certain problem instances, there is no polynomial-time guarantee for these optimizers because nonlinear optimization is generally \texttt{NP}-hard. Often, the problem structure is unknown, and there is no commonly agreed-upon \emph{go-to} optimizer for nonlinear optimization in practice.

Quantum computers are emerging technologies that can leverage the laws of quantum mechanics to offer theoretical and practical advantages over classical computers in solving large-scale computational problems. Unlike their classical counterparts, quantum computers utilize a unique phenomenon known as quantum tunneling to accelerate the solution of nonconvex optimization problems. Specifically, a quantum particle can pass through a high potential barrier that would be insurmountable classically due to insufficient energy. This exotic behavior enables a quantum computer to bypass sub-optimal solutions, efficiently navigating the complex landscape of nonlinear optimization.

Recently, \citet{leng2023quantum} proposes a novel quantum algorithm named Quantum Hamiltonian Descent (QHD). QHD is inspired by the observation that many first-order (i.e., gradient-based) methods can be interpreted as dynamical processes governed by physical laws. For example, it has been shown that the celebrated Nesterov's accelerated gradient descent algorithm can be modeled by a time-dependent Lagrangian mechanical system that would find local minima in the system~\citep{su2016differential,wibisono2016variational}. 
By upgrading the classical Lagrangian mechanics to quantum mechanics, we end up with a minimum-finding quantum process, just like gradient descent. Additionally, this quantum dynamical process demonstrates the quantum tunneling effect, making it a competitive candidate for solving nonconvex optimization problems.
Simulating this quantum dynamical process on a quantum computer gives rise to QHD, a simple but powerful quantum algorithm for continuous optimization, especially nonlinear problems with nonconvex objective functions.
A follow-up work by~\citet{leng2023quantum2} shows that QHD can solve a family of hard optimization instances in polynomial time, while an empirical study suggests that these problem instances are intractable for many classical optimization algorithms such as branch-and-bound, stochastic gradient descent, interior point method, etc.

A key feature of QHD is that it is formulated as a quantum evolution, which can be simulated on both digital and analog quantum computers. This feature allows us to implement QHD to tackle real-world tasks with near-term realizable quantum computers.
Digital quantum computers perform computation by applying a sequence of elementary quantum gates to an initial quantum state. These machines exhibit provable quantum advantages over classical (digital) computers for certain computational tasks, however, they require a large number of digital (i.e., error-corrected) qubits. Although there has recently been a groundbreaking experimental demonstration of early fault tolerance \citep{bluvstein2023logical, google2023suppressing, sivak2023real, singh2023mid}, existing digital quantum computers have not yet reached the size necessary to accelerate the solution of real-world problems in application domains such as management, finance, and engineering \citep{beverland2022assessing,dalzell2023end}.
Analog quantum computers solve computational tasks by configuring and emulating a real quantum system and then performing quantum measurements. These devices are easier to fabricate, control, and scale~\citep{o2009photonic, wendin2017quantum, saffman2016quantum}, while they are unavoidably noisy, and no general error correction technique is currently practical~\citep{lloyd1998analog, atalaya2021continuous}.
\citet{leng2023quantum} proposed a systematic technique named \textit{Hamming encoding} that enables us to implement QHD to solve quadratic programming (QP) problems on analog quantum computers with Ising Hamiltonian.
This technique is exemplified in solving 75-dimensional nonconvex QP problems, where the noisy real-machine implementation of QHD outperforms existing open-source nonlinear optimization software like Ipopt.

In this paper, we develop \qhdopt, an end-to-end implementation of QHD for nonlinear optimization.
A notable feature of \qhdopt\ is that it supports the deployment of QHD to multiple quantum computing hardware, including gate-based quantum computers such as IonQ, and analog quantum computers such as D-Wave.
\qhdopt\ provides a user-friendly interface, with which a nonlinear optimization problem can be specified via either matrix/numeric or symbolic description.
Then, the implementation of QHD is fully automatized and the (approximate) optimal solutions will be returned once the computation is completed.
The mid-level compilation and cross-hardware deployment are achieved by utilizing \simuq\ for Hamiltonian-oriented programming~\citep{peng2023simuq}.

\paragraph{Organization.}
The rest of the paper is organized as follows.\footnote{This paper is not intended to be a comprehensive tutorial or documentation on \qhdopt. Instead, we direct the readers to \url{https://github.com/jiaqileng/QHDOPT} for the source code, examples, tutorials, and documentation.}
In \sec{problem-formulation}, we explain the general problem formulation for nonlinear optimization problems that can be processed and solved by \qhdopt.
In \sec{workflow-main}, we discuss the workflow of \qhdopt\, including the quantum backend and classical refinement.
In \sec{hop}, we discuss several unique design features of \qhdopt, especially the multi-backend compatibility achieved by incorporating the Hamiltonian-oriented programming (HOP) framework.
In \sec{qhd}, we briefly review the QHD algorithm and its implementation on both digital and analog quantum computers.
Then, in \sec{workflow}, we sketch the workflow of the software, including all major steps in the implementation of QHD and classical post-processing. \sec{example} provides two worked examples of modeling and solving nonlinear optimization problems.
In \sec{state}, we review the current state and trend of quantum optimization software.
We conclude with a comparison of \qhdopt\ with other available open-source optimizers in \sec{comparison}.

\subsection{Problem formulation: box-constrained nonlinear optimization}\label{sec:problem-formulation}
The package \texttt{QHDOPT} solves nonlinear programming problems of the following form:

\begin{subequations}\label{eqn:primal}
    \begin{align}
    \min \limits_{x} \ & f(x_1, \dots, x_n) = \underbrace{\sum^n_{i=1}  g_i(x_i)}_{\text{univariate part}} +\underbrace{\sum^{m}_{j=1} p_{j}(x_{k_j})q_{j}(x_{\ell_j})}_{\text{bivariate part}},\label{eqn:obj}\\
    s.t. \ &  L_{i} \le x_i \le U_{i}, \forall i \in \{1,\dots,n\},\label{eqn:box}
\end{align}
\end{subequations}
where $x_1,\dots,x_n$ are $n$ variables subject to the box constraint $x_i \in [L_{i},U_{i}] \subset \mathbb{R}$ for each $i = 1,\dots,n$, and the indices $k_j, \ell_j \in \{1,\dots,n\}$ and $k_j \neq \ell_j$ for each $j = 1,\dots, m$. The functions $g_i(x_{i})$, $p_{j}(x_{k_j})$, and $q_{j}(x_{\ell_j})$ are real univariate differentiable functions defined on $\mathbb{R}$. Note that the univariate part in \eqn{obj} has at most $n$ terms because we can always combine separate univariate functions of a fixed variable $x_i$ into a single one. However, there is no upper bound for the integer $m$ (i.e., the number of bivariate terms).\footnote{It is generally impossible to combine a sum of products into a single product form. For example, we can not find two univariate functions $p(x)$ and $q(y)$ such that $p(x)q(y) = \sin(x)y + xe^y$.}

The nonlinear optimization problem \eqn{primal} is in general NP-hard~\citep{hochbaum2007complexity} and can be used to model several common classes of optimization problems, including linear programming, quadratic programming, and polynomial optimization (with box constraints).
In the following examples, we show how to formulate some standard nonlinear optimization problems in the form of \eqn{obj}.

\begin{example}[Box-constrained Quadratic programming]
    A quadratic programming problem with a box constraint takes the form:
    \begin{subequations}
        \begin{align}
    \min \limits_{x} \ &f(x) \coloneqq \frac{1}{2}x\trans Qx + b\trans x \label{eqn:qp_obj}\\
    s.t. \ & 0 \le x \le 1,
    \end{align}
    \end{subequations}
where $Q\in \R^{n\times n}$ is a symmetric matrix and $ b$ is a real-valued vector of dimension $n$. The objective function can be written as 
$$f(x) = \sum^n_{i = 1} \left(\frac{1}{2}Q_{i,i} x^2_i + b_i x_i\right) + \sum^n_{1 \le k < \ell \le n} Q_{k, \ell }x_k x_{\ell}.$$
This function is represented by \eqn{obj} by choosing 
\begin{subequations}
\begin{align}
&g_i(x_i) = \frac{1}{2}Q_{i,i} x_{i}^2 + b_i x_{i},\quad \forall i = 1,\dots, n, \\
&p_j(x_{k_j}) = Q_{k_j,\ell_j} x_{k_j}, \quad q_j(x_{\ell_j}) = x_{\ell_j}, \quad \forall j \in\left\{1,\dots, \frac{n(n-1)}{2}\right\}.
\end{align}
\end{subequations}
Here $(k_j, \ell_j)$ are the $j$-th pair in the enumeration $\{(k, \ell): 1\leq k<\ell \leq n\}.$
\end{example}
\vspace{4mm}

While the problem formulation can only handle box constraints, we note that many optimization problems with more sophisticated constraints can be reformulated in the form of \eqn{primal} by adding the constraints as a penalty term in the objective function.

\begin{example}[Spherical constraints]
    Consider the optimization problem with $n$ variables:
    \begin{subequations}\label{eqn:spherical}
        \begin{align}
            \min \limits_{x} \ & f(x) \coloneqq \sum^n_{j=1} \alpha_j x_j\\
            s.t. \ &\sum^n_{j=1}x^2_j = 1,
        \end{align}
    \end{subequations}
    where $\alpha_j$ are real scalars for all $j = 1,\dots,n$. The feasible set of this problem is the $n$-dimensional sphere with radius $1$, which can not be directly recast as a box in the form of \eqn{box}. Meanwhile, we observe that all the variables must take values between $0$ and $1$ because the unit sphere is contained in the unit (hyper-)cube. Therefore, we can reformulate \eqn{spherical} to a box-constrained optimization problem by the penalty method:
    \begin{subequations}\label{eqn:penalty}
        \begin{align}
             \min \limits_{x} \ & f(x) \coloneqq \sum^n_{j=1} \alpha_j x_j + \lambda \left(\sum^n_{j=1}x^2_j - 1\right)^2,\label{eqn:penalty-obj}\\
           s.t. \ & 0 \le x \le 1.
        \end{align}
    \end{subequations}
    This new problem can be handled by our software \qhdopt\ since the objective function \eqn{penalty-obj} only involves uni- and bi-variate monomials. As the penalty coefficient $\lambda > 0$ grows, we can show that the solution to the box-constrained problem \eqn{penalty} will eventually converge to the optimal solution to the original problem \eqn{spherical}. 
\end{example}

\vspace{4mm}
It is worth noting that the problem formulation supported by \qhdopt\ is restrictive, and there exist many general nonlinear optimization problems that cannot be directly expressed in~\eqn{primal}. For example, our formulation cannot deal with objective functions involving trivariate monomials (e.g., $xyz$). 
While, in theory, QHD can handle box-constrained optimization models given access to ideal quantum hardware, in \qhdopt\ we limit the appearance of trivariate parts or higher to cater to the current quantum hardware restrictions.

Additionally, we note that there may be several corner cases that are representable by~\eqn{primal} but would require an excessively long time for \qhdopt\ to parse and solve. For example, when the objective function involves thousands of bivariate functions, it might take \qhdopt\ minutes to compile and implement the automatic differentiation subroutine based on JAX. We advise users to prioritize the use cases with low-degree polynomials, bounded exponential functions, and simple trigonometric functions.

\subsection{Solving problems in \qhdopt}\label{sec:workflow-main}
\qhdopt\ utilizes the Quantum Hamiltonian Descent algorithm to facilitate the solution of nonlinear and nonconvex optimization problems.
Theoretically, Quantum Hamiltonian Descent, when running with an ideal fault-tolerant quantum computer, can solve many optimization problems up to global optimality given sufficiently long runtime~\cite{leng2023quantum}.
However, at the current stage, due to the lack of fault tolerance, we can only implement Quantum Hamiltonian Descent in a \textit{low-precision} and \textit{noisy} manner, which significantly reduces the solution quality promised by the theoretical guarantee.
To mitigate the noisy performance of near-term quantum hardware with limited resources, we adopt a hybrid quantum-classical computing workflow in \qhdopt\ to achieve optimal performance, as illustrated in~\fig{workflow}B.

\paragraph{Pre-processing and problem encoding.}
First, we map a box-constrained nonlinear optimization problem to a quantum-mechanical system with finite degrees of freedom. This reduced quantum model can be regarded as a finite-precision approximation of the original QHD model.
Then, this quantum model is embedded into a larger quantum system that is natively executable using one of the supported quantum backends. This process is called \textit{Hamiltonian programming}. While the quantum hardware only ``sees'' a reduced version of the original problem, the \textit{Hamiltonian embedding} technique~\cite{leng2024expanding} ensures that the spatial structure inherited from the original problem is preserved and naturally encoded in the quantum operator. Therefore, \qhdopt\ allows us to run a coarse-grained version of QHD on near-term quantum devices.

\paragraph{Deployment and decoding.}
Then, the quantum operator that encodes the original nonlinear optimization problem is constructed and executed on a quantum backend. Currently, \qhdopt\ supports three backends: the D-Wave quantum computer, the IonQ quantum computer, and a classical simulator based on QuTiP. The measurement results from quantum devices are in 0-1 format (i.e., binaries), which requires a decoder to recover the corresponding solution in the continuous space (e.g., the unit box). 

\paragraph{Classical refinement.}
Limited by the size and coherent time of current quantum devices, the quantum-generated solutions are of low precision and intrinsically noisy. \qhdopt\ relies on classical local search algorithms, such as first- and second-order methods, to improve numerical precision. Currently, \qhdopt\ supports two local optimizers: a general-purpose interior point method (Ipopt) and a truncated Newton method (TNC) implemented in SciPy. While we do not include other local search subroutines in \qhdopt, we note that generic local optimizers allowing box constraints should work as well.

\vspace{4mm}
Since \qhdopt\ leverages local search algorithms as refiners, the output solutions are necessarily locally optimal (i.e., first- or second-order stationary points, depending on the choice of refinement subroutine).
That being said, we would like to note that the Quantum Hamiltonian Descent (QHD) algorithm, when executed on a large fault-tolerant quantum computer, is able to find the global minimum for a large family of nonconvex functions with mild assumptions, provided that the runtime is sufficiently long \citep[Theorem 2]{leng2023quantum}. The performance of \qhdopt\ for practical problems, however, heavily depends on the quality of near-term quantum devices, which are often of limited scale and prone to physical noise. Meanwhile, it is also possible to refine the quantum-generated solutions using a global solver (e.g., Gurobi, BARON); in this case, the global optimality is guaranteed, but the post-processing time could be significantly longer. Due to the limited time frame, we leave a global-solver-based refinement as future work.

\subsection{Unique design features}\label{sec:hop}

\begin{figure*}
    \centering
    \includegraphics[width=0.95\linewidth, trim = 2cm 3.5cm 5.5cm 3.5cm, clip]{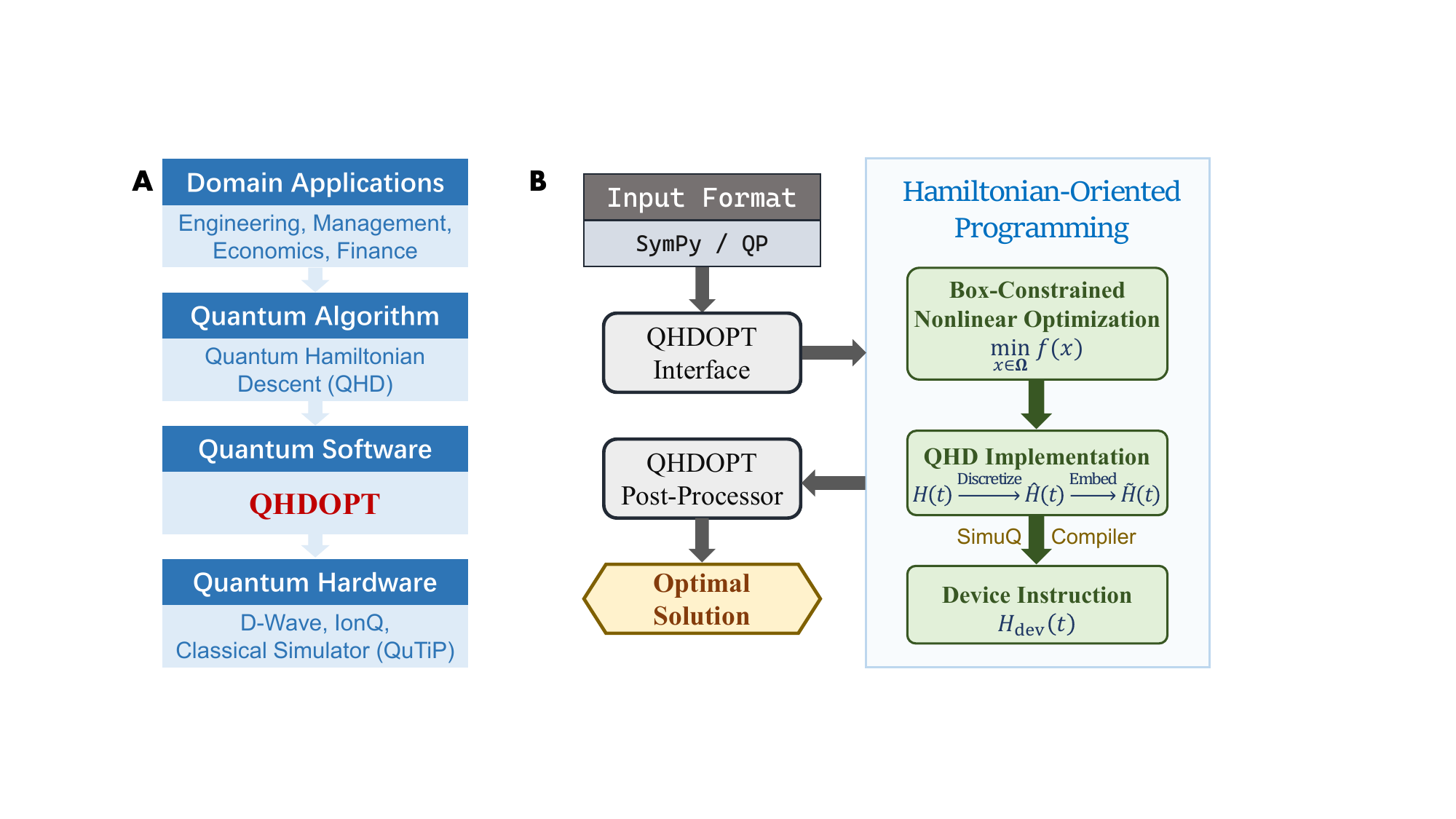}
    \caption{\small An Overview of \qhdopt.~\textbf{A.} Building the stack of quantum computing for nonlinear optimization. \textbf{B.} The workflow of \qhdopt,~inspired by the Hamiltonian-oriented programming paradigm.}
    \label{fig:workflow}
\end{figure*}

In what follows, we discuss a few unique design features of our software.

\paragraph{Hamiltonian-oriented programming (HOP).}
\qhdopt\ exploits the Quantum Hamiltonian Descent (QHD) algorithm to solve nonlinear optimization problems. This quantum algorithm is formulated as a Hamiltonian simulation (i.e., simulating the evolution of a quantum-mechanical system), encompassing a novel abstraction of computation on quantum devices, which we call \textbf{Hamiltonian-Oriented Programming (HOP)}.
In contrast to the conventional circuit-based quantum computation paradigm where theorists describe quantum algorithms in terms of quantum circuits, the HOP paradigm describes quantum algorithms as a single or a sequence of quantum Hamiltonian evolution. 
This new paradigm enables us to build a stack of quantum applications by leveraging the native programmability of quantum hardware in the development of quantum algorithms and software, as illustrated in \fig{workflow}A. 
The HOP paradigm is empowered by \simuq, a recent framework for programming and compiling quantum Hamiltonian systems by \citet{peng2023simuq}. 
In \simuq,~the programming and simulation of quantum Hamiltonian systems are wrapped in user-friendly Python methods. This makes the high-level programming and deployment of Hamiltonian-oriented quantum algorithms accessible to users with little exposure to real-machine engineering and manipulation.
A detailed discussion on the Hamiltonian programming and compilation in \qhdopt\ is available in \sec{workflow}.

\paragraph{Multi-backend compatibility.}
In \qhdopt, we utilize \simuq\ as an intermediate layer for the programming of QHD and leverage the \simuq\ compiler to realize multi-backend compatibility. Through \simuq, \qhdopt\ initially constructs a hardware-agnostic Hamiltonian representation of QHD (i.e., Hamiltonian embedding) that can be deployed on various quantum backends, including D-Wave devices, IonQ devices, and classical simulators via QuTiP \citep{johansson2012qutip}.

\paragraph{Automatic differentiation.}
\qhdopt\ relies on JAX, a high-performance numerical computing library, to perform automatic differentiation of smooth, nonlinear objective functions. This feature enables \qhdopt\ to seamlessly post-process quantum-generated solutions using local search optimizers.

\section{Quantum Hamiltonian Descent}\label{sec:qhd}
In our software, we utilize QHD to solve box-constrained nonlinear optimization problems as described in \eqn{primal}. QHD solves a continuous optimization problem by simulating a quantum dynamical system governed by an evolutionary partial differential equation called \emph{Schr\"odinger equation}. Here, we give a high-level review of this quantum algorithm and more details can be found in \citep{leng2023quantum}. 

\subsection{Mathematical formulation and interpretation}\label{sec:math-formulation}
Consider a nonlinear objective function $f(x)$ with a box constraint $\Omega= \{(x_1,\dots,x_n)\in \R^n\colon L_i \le x_i \le U_i, \forall i = 1,\dots, n\}$. To solve this optimization problem, QHD requires simulating the following Schr\"odinger equation over the feasible set $\Omega$ with Dirichlet boundary condition, i.e., $\Psi(t,x) = 0$ for $x \in \partial \Omega$,
\begin{equation}\label{eqn:qhd-pde}
    i \frac{\partial}{\partial t}\Psi(t,x) = \left[e^{\varphi_t}\left(-\frac{1}{2}\Delta \right) + e^{\chi_t}f(x)\right]\Psi(t,x),
\end{equation}
subject to an initial state $\Psi(t,x) = \Psi_0(x)$. Here, the operator $\Delta\coloneqq \sum^n_{i=1}\frac{\partial^2}{\partial x^2_i}$ is the Laplacian operator defined in the interior of $\Omega$, and the time-dependent functions $e^{\varphi_t}$ and $e^{\chi_t}$ control the total energy distribution of the quantum system. In practice, the initial state $\Psi_0(x)$ is often chosen as a quantum state that is easy to prepare, for example, a Gaussian state or a uniformly random state. For general (nonconvex) optimization problems, it is observed that an inverse polynomially decaying $e^{\varphi_t}$ and polynomially increasing $e^{\chi_t}$ (e.g., $\varphi_t = -\log(1+\gamma t^2)$, $\chi_t = \log(1+\gamma t^2)$ with a positive $\gamma$) work well for many test problems~\cite{leng2023quantum}. With a Gaussian initial state and smooth time-dependent functions, the dynamics generated by~\eqn{qhd-pde} can be simulated using $\widetilde{\mathcal{O}}(nT)$ elementary gates and $\widetilde{\mathcal{O}}(T)$ queries to the objective function $f$~\cite{childs2022quantum}.

Physically, the equation~\eqn{qhd-pde} describes the time evolution of a quantum particle in the box $\Omega$. The time-dependent functions $e^{\varphi_t}$ and $e^{\chi_t}$ control the total energy distribution of this quantum particle: when their ratio $e^{\varphi_t}/e^{\chi_t}$ is large, the kinetic energy dominates and the particle tends to bounce around; otherwise, the potential energy takes over and the particle tends to stay still. If we choose these functions such that $\lim_{t\to \infty} e^{\varphi_t}/e^{\chi_t} = 0$, the kinetic energy of the system is dissipated over time and eventually the quantum particle will take a low-energy configuration. At this point, if we measure this quantum particle, the measured position (which must lie in the feasible set $\Omega$) is likely to give an approximate solution to the problem $f(x)$. In some sense, QHD can be regarded as a quantum version of Polyak's heavy ball method~\citep{polyak1964some,attouch2000heavy}.

QHD describes a quantum particle exploring the optimization landscape $f(x)$. When a high-energy barrier emerges, the quantum particle may leverage the quantum tunneling effect to go through the barrier and find a lower local minimum. However, simulating the quantum evolution \eqn{qhd-pde} with a classical computer would require exponential space and time, making this idea impractical as a classical optimization algorithm. 
On the other hand, the evolution~\eqn{qhd-pde} can be efficiently simulated using a quantum computer, which makes QHD a genuine quantum algorithm that can leverage the quantum tunneling effect for nonconvex optimization.
Theoretically, it has been shown that QHD can efficiently find the global minimum for certain nonconvex problems with exponentially many local minima, while many classical optimizers such as simulated annealing and SGD appear to require a much longer time to obtain a global solution~\citep{leng2023quantum2}. Numerical experiments also show that QHD outperforms classical first- and second-order methods in a broad class of nonconvex problems with many local stationary points~\citep{leng2023quantum}.

\subsection{Real-machine implementation}

Quantum Hamiltonian Descent is formulated as a \textit{Hamiltonian simulation} task, i.e., solving a quantum Schrödinger equation as in~\eqn{qhd-pde}. While efficient quantum algorithms, such as those proposed by~\cite{childs2022quantum}, can tackle this simulation task exponentially faster than any known classical algorithms, these quantum simulation algorithms require large fault-tolerant quantum computers. Such ideal quantum computing hardware has not yet been realized due to the immature progress of quantum technology.

To fully exploit the limited programmability of current quantum hardware such as D-Wave and IonQ, \qhdopt\ employs a technique named \emph{Hamiltonian embedding}~\citep{leng2024expanding} to implement QHD. This technique enables us to map the QHD Hamiltonian to a larger Hamiltonian, and the latter can be \textit{natively} simulated on existing quantum devices. This real-machine implementation technique is detailed in \sec{ham-program-compile}.

\section{The Workflow of \qhdopt}\label{sec:workflow}

\subsection{Modeling of nonlinear problems}
\label{sec:model}

\qhdopt\ offers support for two Python-based input formats: the SymPy format for \emph{symbolic} input and the QP format for \emph{numerical} input (i.e., arrays). These two input formats enable users to define their target optimization problems both efficiently and with great flexibility.

\begin{figure*}[!ht]
    \centering
    \begin{adjustbox}{minipage=0.45\linewidth, scale=0.85}
\hspace{1em}
    \begin{subfigure}[t]{\linewidth}
    \begin{pythonnum}
from qhdopt import QHD
from sympy import symbols, exp

x, y = symbols("x y")
f = y**1.5 - (y-0.75) * exp(4*x)
model = QHD.SymPy(f, [x, y])
    \end{pythonnum}
    \caption{An example using the SymPy input format}
    \label{fig:sympy-exp}
    \end{subfigure}
    \end{adjustbox}
    \qquad \qquad
    \begin{adjustbox}{minipage=0.50\linewidth, scale=0.85}
    \begin{subfigure}[t]{\linewidth}
    \begin{pythonnum}
from qhdopt import QHD

Q = [[-8, 3],
     [3, -4]]
b = [3, -1]
model = QHD.QP(Q, b)
    \end{pythonnum}
    \caption{An example using the QP input format}
    \label{fig:qp-exp}
    \end{subfigure}
    \end{adjustbox}
    \caption{\small Input formats in \qhdopt. \qhdopt\ supports both symbolic and numerical input formats.}
    \label{fig:python-code}
\end{figure*}

\paragraph{SymPy format.} \texttt{SymPy} \citep{meurer2017sympy} is a Python package that supports symbolic expression processes. Users can specify $f(x)$ in \eqn{primal} by declaring variables in \texttt{SymPy} and constructing the expression, as in the code snippet in \fig{sympy-exp}. Here, we import necessary functions like $\exp$ from \texttt{SymPy} and \pyth{QHD} from package \qhdopt\ in lines 1 and 2. We declare variables \pyth{x} and \pyth{y} in \texttt{SymPy} using \pyth{symbols} in line 3 where the passed string is for \texttt{SymPy} to print the expressions. In line 4, we construct the function $f(x)$, where \pyth{y**1.5} represents the exponential $y^{1.5}$, \pyth{exp(4*x)} represents $e^{4x}$, and so on. Lastly, we create a QHD model instance in line 5 and pass \pyth{f} and a symbol list \pyth{[x, y]} to it, informing the QHD model the target optimization function is $f$ with symbols $x$ and $y$.

\paragraph{QP format.} For users with specific interests in quadratic programming (QP), we provide a more efficient input model for them. To specify a QP instance with objective function $f(x) = \frac{1}{2}x\trans Q x + b\trans x$, we can directly input the matrices $Q$ and $b$, as in the code snippet in \fig{qp-exp}. First, we construct $Q$ by a nested Python list or a NumPy array in lines 2 and 3. It is required that $Q$ forms a symmetric square matrix. Then we input the vector $b$ as \pyth{b}. Similar to \texttt{SymPy}, we construct the instance by calling the \pyth{QP} method from \qhdopt\ and pass \pyth{Q, b} into it.

\subsection{Hamiltonian programming and compilation}\label{sec:ham-program-compile}
Once a nonlinear optimization problem $f(x)$ is defined using one of the supported input formats, \texttt{QHDOPT} will form a Hamiltonian description of the corresponding QHD algorithm, as described in \eqn{qhd-pde}. This Hamiltonian description serves as an intermediate layer in the compilation stack and is independent of the choice of the backend (i.e., hardware-agnostic). Although \texttt{QHDOPT} automates this process, making manual execution unnecessary in most cases, we provide detailed discussions for readers who are interested in gaining a deeper understanding of our software's design.

There are two major steps in the construction of the Hamiltonian description of QHD, namely, \emph{spatial discretization} and \emph{Hamiltonian embedding}.

\subsubsection{Spatial discretization}
First, we need to perform spatial discretization of the QHD Hamiltonian (which is an unbounded operator) so that it can be described by a finite-dimensional quantum system. For a thorough and mathematically rigorous discussion, readers are encouraged to refer to \cite[Appendix F.2.1]{leng2023quantum}. Given a nonlinear optimization in the form of \eqn{primal}, the QHD Hamiltonian reads the following,
$$H(t) = e^{\varphi_t}\left(-\frac{1}{2}\Delta \right) + e^{\chi_t}\left(\sum^n_{i=1}  g_i(x_i) +\sum^{m}_{j=1} p_{j}(x_{k_j})q_{j}(x_{\ell_j})\right),$$
which acts on any $L^2$-integrable functions over the feasible set $\Omega = \{(x_1,\dots,x_n)\in \R^n\colon L_i \le x_i \le {U_i}, \forall i = 1,\dots, n\}$. Here, for simplicity, we assume the feasible set is the unit box, i.e., $L_i = 0$ and ${U_i} = 1$ for all $i = 1,\dots,n$.
We utilize the centered finite difference scheme to discretize this differential operator. Suppose that we divide each dimension of the unit box $\Omega$ using $N$ quadrature points $\left\{0, h,\dots, (N-2)h, 1\right\}$ (where $h = 1/(N-1)$), the resulting discretized QHD Hamiltonian is an $N^n$-dimensional operator of the form,
\begin{align}\label{eqn:discretized_qhd}
    \hat{H}(t) = e^{\varphi_t}\left(-\frac{1}{2}L_d\right) + e^{\chi_t}F_d,
\end{align}
where (assuming $k_j < \ell_j$ for all $j = 1,\dots,m$)
$$L_d = \sum^n_{i=1} I\otimes \dots \otimes \underbrace{L}_{\text{the $i$-th operator}} \otimes \dots I,$$
$$F_d = \sum^n_{i=1} I\otimes \dots \otimes \underbrace{D(g_i)}_{\text{the $i$-th operator}} \otimes \dots I + \sum^m_{j=1} I\otimes \dots \otimes \underbrace{D(p_j)}_{\text{the $k_j$-th operator}} \otimes \dots \otimes \underbrace{D(q_j)}_{\text{the $\ell_j$-th operator}} \otimes \dots I.$$
Here, $I$ is the $N$-dimensional identity operator, $L$ and $D(g)$ are $N$-dimensional matrices given by ($g$ is a differentiable function defined on $[0,1]$ and $g_i \coloneqq g(ih)$ for $i = 0,\dots,N-1$),
$$L = \frac{1}{h^2} \begin{bmatrix} -2 & 1 & & \\1 & -2 & 1 & \\...& ... & ... &...\\& 1 & -2 & 1\\& & 1 &-2\\\end{bmatrix},\quad D(g) = \begin{bmatrix} g_0 & & & \\ & g_1 &  & \\...& ... & ... &...\\&  & g_{N-2}& \\& & &g_{N-1}\\\end{bmatrix}.$$
The tridiagonal $L$ matrix corresponds to the finite difference discretization of the second-order differential operator $\frac{d^2}{d x^2}$, and the diagonal matrix $D(g)$ corresponds to the finite difference discretization of the univariate function $g(x)$. Note that $L$ has a global phase $-2/h^2$, i.e., $L = L' - 2/h^2$ with $L'$ only contains the off-diagonal part of $L$. Since the global phase does not affect the quantum evolution (therefore, the result of the QHD algorithm), we replace $L$ with $L'$ in the rest of the discussion.

\subsubsection{Hamiltonian embedding}
The discretized QHD Hamiltonian, as described in \eqn{discretized_qhd}, is a Hermitian matrix with an explicit tensor product decomposition structure. This particular structure allows us to leverage the Hamiltonian embedding technique~\citep{leng2024expanding} to construct a surrogate Hamiltonian $\Tilde{H}(t)$ such that the QHD algorithm (i.e., simulating the Hamiltonian $\hat{H}(t)$) can be executed by simulating $\Tilde{H}(t)$. In our case, the surrogate Hamiltonian $\Tilde{H}(t)$ is an Ising-type quantum Hamiltonian that involves at most $nN$ qubits and $\max(n,m) N$ two-body interaction terms. This means $\Tilde{H}(t)$ can be efficiently simulated on current quantum computers, including IonQ's trapped ion systems and D-Wave's quantum annealer.

To construct the Hamiltonian embedding of $\hat{H}(t)$, the first step is to build the Hamiltonian embeddings of the $N$-by-$N$ matrices $L'$ and $D(g)$ (for arbitrary differentiable function $g$). Both are sparse matrices so we can utilize the embedding schemes provided in \cite[Section~2.3]{leng2024expanding}. \qhdopt\ allows users to choose from three embedding schemes: Hamming\footnote{The details of Hamming embedding can be found in \cite[Appendix F.3]{leng2023quantum}. Note that this embedding scheme is referred to as ``Hamming encoding'' in \citep{{leng2023quantum}}.}, unary, and one-hot\footnote{More precisely, the one-hot embedding we implemented in \texttt{QHDOPT} is referred to as ``penalty-free one-hot'' embedding in \citep{leng2024expanding}.}. In \tab{embedding-scheme}, we list the details of these embedding schemes when applied to $L'$ and $D(g)$. 
We note that the Hamming embedding scheme only works for quadratic programming, while the other two schemes (unary, one-hot) work for a broader class of nonlinear functions such as exponential functions.
To be consistent with our source code, we adopt the left-to-right $0$-indexing system for bits/qubits, e.g., $1_0 0_1 1_2 1_3$.

\begin{table}[!ht]
    \centering
    \begin{adjustbox}{width = 0.97\linewidth}
    \begin{tabular}{|p{2cm}|p{2.5cm}|p{2.5cm}|p{2cm}|p{4cm}|p{4cm}|}
     \hline
     \textbf{Embedding scheme} & \textbf{Supported input format} & \textbf{Supported backend} & \textbf{Number of qubits} & \textbf{Embedding of $L'$} & \textbf{Embedding of $D(g)$}\\
     \hline
     Hamming & QP, SymPy & QuTiP, IonQ, D-Wave & $r = N-1$ & $\sum^{r-1}_{k=0}\mathbf{X}_k/h^2$ & Only for QP, see discussions below\\
     \hline
     Unary & QP, SymPy & QuTiP, IonQ, D-Wave & $r = N-1$ & $\sum^{r-1}_{k=0}\mathbf{X}_k/h^2$ & $\sum^{r-1}_{k=0} (g_{r-k}-g_{r-k-1})\mathbf{n}_k + g_0 \mathbf{I}$\\
     \hline
     One-hot & QP, SymPy & QuTiP, IonQ & $r = N$ & $\sum^{r-2}_{k=0}(\mathbf{X}_{k}\mathbf{X}_{k+1} + \mathbf{Y}_{k}\mathbf{Y}_{k+1})/(2h^2)$ & $\sum^{r-1}_{k=0}g_{r-1-k} \mathbf{n}_{k}$\\
     \hline
    \end{tabular}
    \end{adjustbox}
    \caption{Embedding schemes supported by \texttt{QHDOPT}}
    \label{tab:embedding-scheme}
\end{table}

In \tab{embedding-scheme}, the integer $r$ represents the number of qubits used to embed an $N$-dimensional matrix. The operators $\mathbf{X}_k$, $\mathbf{Y}_k$, and $\mathbf{n}_k$ are the Pauli-X, Pauli-Y, and number operator acting at site $k$, respectively,
$$\mathbf{X} = \begin{bmatrix}0 & 1\\1 & 0\end{bmatrix},\quad \mathbf{Y} = \begin{bmatrix}0 & -i\\i & 0\end{bmatrix},\quad \mathbf{n} = \begin{bmatrix}0 & 0\\0 & 1\end{bmatrix}.$$
Since the Hamming embedding scheme is only allowed for quadratic programming, we do not consider the Hamming embedding for general nonlinear functions $g$. Instead, we only consider the embedding of the identity and quadratic functions, i.e., $g(x) = x$ and $g(x) = x^2$, their corresponding Hamming embeddings are $\mathcal{E}_1 = \frac{1}{r}\sum^{r-1}_{k=0} \mathbf{n}_k$ and $\mathcal{E}_2 = (\mathcal{E}_1)^2$, respectively.

Now, we denote $\mathcal{E}^{(i)}[A]$ as a Hamiltonian embedding of an $N$-by-$N$ Hermitian matrix $A$ acting on sites $(i-1)r,(i-1)r+1,\dots,ir-1$, where $i = 1,\dots,n$. Then, using the rules of building Hamiltonian embeddings~\cite[Theorem 2]{leng2024expanding}, we obtain a $nr$-qubit Hamiltonian that embeds the discretized QHD Hamiltonian $\hat{H}(t)$,
\begin{align}
    \Tilde{H}(t) = e^{\varphi_t} \left(-\frac{1}{2}\sum^{n}_{i=1}\mathcal{E}^{(i)}[L']\right) + e^{\chi_t}\left(\sum^{n}_{i=1}\mathcal{E}^{(i)}[D(g_i)] + \sum^{m}_{j=1}\mathcal{E}^{(k_j)}[D(p_j)]\mathcal{E}^{(\ell_j)}[D(q_j)]\right). 
\end{align}

\begin{example}[One-hot embedding for $x_1x_2$]
    We give a simple example for the Hamiltonian embedding of the discretized QHD Hamiltonian when the objective function is $f(x_1,x_2) = x_1x_2$. This objective only involves a single bivariate term with $p(x) = q(x) = x$. We use the one-hot embedding with $N = r = 3$. Then, the Hamiltonian embeddings of $L'$ and $D(x)$ are,
    \begin{align*}
        \mathcal{E}[L'] = \frac{1}{2h^2}\left(\mathbf{X}_0\mathbf{X}_1 + \mathbf{X}_1\mathbf{X}_2 + \mathbf{Y}_0\mathbf{Y}_1 + \mathbf{Y}_1\mathbf{Y}_2\right),\quad
        \mathcal{E}[D(x)] = \mathbf{n}_0 + \frac{1}{2}\mathbf{n}_1,
    \end{align*}
    respectively. As a result, the full Hamiltonian embedding reads
    \begin{align*}
        \Tilde{H}(t) = &\frac{2e^{\varphi_t}}{h^2}\left(\mathbf{X}_0\mathbf{X}_1 + \mathbf{X}_1\mathbf{X}_2 + \mathbf{X}_3\mathbf{X}_4 + \mathbf{X}_4\mathbf{X}_5 + \mathbf{Y}_0\mathbf{Y}_1 + \mathbf{Y}_1\mathbf{Y}_2 + \mathbf{Y}_3\mathbf{Y}_4 + \mathbf{Y}_4\mathbf{Y}_5\right) + \\
        &e^{\chi_t} \left(\mathbf{n}_0 + \frac{1}{2}\mathbf{n}_1\right)\left(\mathbf{n}_3 + \frac{1}{2}\mathbf{n}_4\right).
    \end{align*}
\end{example}

In \qhdopt, we use \simuq\ to construct the Hamiltonian embedding $\Tilde{H}(t)$. The users only need to specify the number of qubits $r$ (for each continuous variable), the embedding scheme, and a desired backend in the \pyth{model.optimize()} function, as detailed in the next subsection.

\subsection{Deployment and post-processing}

When the Hamiltonian embedding $\Tilde{H}(t)$ of a given problem is built, it can be executed on a supported quantum backend by running \pyth{optimize()} (refer to \sec{example} for sample code). The quantum measurement results are then retrieved from the executing backend in the form of bitstrings. Following this, \qhdopt\ implements a series of classical post-processing subroutines. These include decoding the raw measurement results (i.e., bitstrings) into low-resolution solutions and refining them via a classical local solver. The refined solutions are then returned to the users as final results.

\subsubsection{Deployment on quantum devices}
Currently, \qhdopt\ supports three backend devices for deployment, including classical simulators (e.g., QuTiP), IonQ, and D-Wave. For all three backend devices, the quantum register is initialized to the uniform superposition state. On the IonQ device, the uniform superposition state can be prepared using a single layer of Hadamard gate; on the D-Wave device, the uniform superposition state is the default initial state and it can be prepared in microseconds. When deployed on IonQ, \qhdopt\ uses $\varphi_t = -\log(1+\gamma t^2)$ and $\chi_t = \log(1+\gamma t^2)$ for the time-dependent functions (see~\sec{math-formulation} for details).
The time-dependent functions on D-Wave are more restricted and they can only be specified as piece-wise linear functions. We find the default annealing schedule (20 microseconds) provided by the D-Wave device usually works well in practice. We also showcase user-specified time-dependent functions (annealing schedules) in a notebook in the ``examples'' folder.

Here, we demonstrate the deployment procedures in \qhdopt\ using the D-Wave backend, while the same process applies to the other two backends. 
In \fig{dwave-execute}, a snippet of the source code for the function \pyth{QHD.dwave_exec()} is displayed. After programming the Hamiltonian embedding and the quantum system realizing QHD (lines 2-3), we initiate an abstract D-Wave machine (line 5). Then \qhdopt\ employ \simuq\ to compile the Hamiltonian embedding into low-level device instructions readable by D-Wave (line 7) using \simuq's \pyth{DWaveProvider()}, effectively generating Hamiltonian $H_{\mathrm{dev}}(t)$ on the D-Wave devices. Next, the instructions are sent to the D-Wave quantum computer to execute (line 9), and the raw quantum samples are collected by \qhdopt\ as bitstrings (line 12-17).

\begin{figure*}[!ht]
    \centering
    \begin{adjustbox}{minipage=1\linewidth, scale=0.9}
    \hspace{1em}
    \begin{pythonnum}
def dwave_exec(self, verbose=0):
    Hamiltonian, T = ...
    self.qs.add_evolution(Hamiltonian, T) # Hamiltonian embedding
    
    dwp = DWaveProvider(self.api_key) # initiate D-Wave machine
    ...
    dwp.compile(self.qs, self.shots)
    ...
    dwp.run()
    ... 

    self.raw_result = dwp.results()
    raw_samples = []
    for i in range(self.shots):
        raw_samples.append(QHD.spin_to_bitstring(self.raw_result[i]))
    
    return raw_samples
    \end{pythonnum}
    \end{adjustbox}
    \caption{\small Deploying and executing QHD on the D-Wave quantum computer}
    \label{fig:dwave-execute}
\end{figure*}

\subsubsection{Decoding}
As we have seen, the real-machine results are in the bitstring format because they are retrieved by computational basis measurements in the quantum computer.
These bitstrings need to be converted to floating-point arrays via the built-in decoder, as presented in \fig{decoder}. This decoder maps a bitstring to a floating-point array that represents a low-resolution solution to the input optimization problem. For example, if we use the unary embedding for a 2-dimensional problem with resolution parameter $r = 4$, the decoder will map an 8-bit string to a length-2 array. For example, 00010011 is mapped to $[0.25, 0.5]$. More details of the embedding schemes and their decoding are available in \citep{leng2024expanding}.

\begin{figure*}[!ht]
    \centering
    \begin{adjustbox}{minipage=1\linewidth, scale=0.9}
    \hspace{1em}
    \begin{pythonnum}
def bitstring_to_vec(self, bitstring, d, r):
    if self.embedding_scheme == 'unary':
        return QHD.unary_bitstring_to_vec(bitstring, d, r)
    elif self.embedding_scheme == 'onehot':
        return QHD.onehot_bitstring_to_vec(bitstring, d, r)
    elif self.embedding_scheme == 'hamming':
        return QHD.hamming_bitstring_to_vec(bitstring, d, r)
    else:
        raise Exception("Illegal embedding scheme.")
    \end{pythonnum}
    \end{adjustbox}
    \caption{\small Bitstring-to-vector decoder in \qhdopt}
    \label{fig:decoder}
\end{figure*}

\subsubsection{Refinement}
Limited by the size of current quantum devices, in most cases, we can only use a small resolution parameter (e.g., $r =8$) in the real-machine implementation of QHD. Therefore, the retrieved measurement results are merely low-resolution solutions to the specified optimization problem. To improve the precision of the solutions, \qhdopt\ then post-processes the measurement results using local search classical optimization methods. 

In principle, any generic local optimizers allowing box constraints should work as well; due to the limited resources, we provide two classical refinement options for the users in \qhdopt, including the truncated Newton method (TNC) using SciPy and the interior point method using Ipopt. TNC is a quasi-Newton method, and the interior point method exploited by Ipopt is a second-order method. 
These classical refiners require the gradient and/or Hessian information of the objective functions. For quadratic programming problems (using the QP input format), the gradient and Hessian can be computed explicitly:
$$\nabla f(x) = Q x,\quad H f(x) = Q.$$
For more general nonlinear optimization problems specified using the SymPy input format, \qhdopt\ computes the gradient and Hessian information by employing Jax~\citep{frostig2018compiling}, a high-performance numerical computing library developed by Google, to perform auto-differentiation.

The post-processed results can be retrieved from \pyth{model.post_processed_samples}. By default, the post-processing subroutine is enabled and automatically executed by running \pyth{optimize()}. However, users can also disable post-processing by specifying \pyth{optimize(fine_tune=False)}. In this case, \pyth{model.post_processed_samples} returns \pyth{None} type.

\section{Examples using \qhdopt}\label{sec:example}

In this section, we exhibit two simple examples showcasing the use cases of \qhdopt.

\subsection{Quadratic programming}

We first consider a 2-dimensional quadratic programming problem, whose objective function is defined as follows,
\begin{align}
    f(x,y) = -x^2 + xy - \frac{1}{2}y^2 + \frac{3}{4}x - \frac{1}{4}y =\frac{1}{2}\begin{bmatrix} x & y \end{bmatrix} \begin{bmatrix} -2 & 1 \\ 1 & -1\end{bmatrix}\begin{bmatrix} x \\ y \end{bmatrix}+\begin{bmatrix} \frac{3}{4} & -\frac{1}{4}\end{bmatrix}\begin{bmatrix} x \\ y \end{bmatrix},
\end{align}
for $x, y \in [0,1].$

In \fig{example-1}, we exemplify using \qhdopt\ to solve this QP problem with all three backends. Note that the API key (not included in \qhdopt) is required to access cloud-based quantum computers such as D-Wave and IonQ.
By default, \pyth{QHD.optimize()} automatically executes the Scipy TNC method to fine-tune the raw quantum measurement data. To switch to the Ipopt optimizer in the fine-tuning step, one can specify \pyth{post_processing_method="IPOPT"} in the setup, as shown in line 17.

\begin{figure*}[!ht]
    \centering
    \begin{adjustbox}{minipage=1\linewidth, scale=0.9}
    \hspace{1em}
    \begin{pythonnum}
from qhdopt import QHD

# Using QP input format
Q = [[-2, 1],[1, -1]]
b = [3/4, -1/4]
model = QHD.QP(Q, b, bounds=(0,1))

# Deployment # 1: D-Wave (default embedding scheme: unary)
model.dwave_setup(8, api_key="DWAVE_API_KEY")
model.optimize()

# Deployment # 2: IonQ (default embedding scheme: one-hot)
model.ionq_setup(6, api_key="IONQ_API_KEY", time_discretization=30)
model.optimize()

# Model 3: QuTiP simulator (default embedding scheme: one-hot)
model.qutip_setup(6, post_processing_method="IPOPT")
model.optimize()
    \end{pythonnum}
    \end{adjustbox}
    \caption{\small Solving a quadratic programming problem using \qhdopt}
    \label{fig:example-1}
\end{figure*}

\begin{figure*}[!ht]
    \centering
    \begin{adjustbox}{minipage=0.5\linewidth, scale=0.85}
\hspace{1em}
    \begin{subfigure}[t]{\linewidth}
    \begin{pythonnum}
from sympy import symbols, exp
from qhdopt import QHD

# Using SymPy input format
x, y = symbols("x y")
f = y**1.5 - exp(4*x) * (y-0.75)
model = QHD.SymPy(f, [x, y], 
                  bounds=(0,1))

# Deploying QHD on the D-Wave device
model.dwave_setup(8, api_key="API_KEY")
model.optimize(verbose=1)
    \end{pythonnum}
    \caption{Deploying QHD on D-Wave}
    \label{fig:example-2}
    \end{subfigure}
    \end{adjustbox}
    \qquad \qquad
    \begin{adjustbox}{minipage=0.40\linewidth, scale=0.73}
    \begin{subfigure}[t]{\linewidth}
    \begin{pythonnum}
* Coarse solution
    Minimizer: [1. 0. 1.]
    Minimum: -1.0
* Fine-tuned solution
    Minimizer: [1. 0. 1.]
    Minimum: -1.0
    Success rate: 1.0
* Runtime breakdown
    SimuQ compilation: 0.000 s
    Backend QPU runtime: 0.119 s
    Backend overhead time: 3.825 s
    Decoding time: 0.019 s
    Fine-tuning time: 0.161 s
* Total time: 4.124 s
    \end{pythonnum}
    \caption{Execution Summary}
    \label{fig:example2-readout}
    \end{subfigure}
    \end{adjustbox}
    \caption{\small Solving a nonlinear optimization problem using \qhdopt. The backend overhead time includes network transmission time, queue time, etc.}
\end{figure*}

\subsection{Nonlinear optimization involving exponential function}

Next, we consider the following nonlinear minimization problem with the objective function:
\begin{align}
    f(x,y) = y^{3/2} - e^{4x}\left(y-\frac{3}{4}\right),\quad x,y\in [0,1].
\end{align}
This objective function $f(x,y)$ is not a polynomial; it involves a fractional power and an exponential function.
In \fig{example-2}, we illustrate a sample code that runs \qhdopt\ to solve the problem defined above. The function is constructed using the SymPy input format, then deployed on the D-Wave quantum computer with resolution parameter $r=8$.
We may set \pyth{verbose=1} to print a detailed summary of this execution, including the best-so-far coarse and fine-tuned solutions, as well as a total runtime breakdown, as shown in \fig{example2-readout}.

\section{The State of Software for Quantum Optimization}\label{sec:state}

Software packages are crucial for lowering the barrier to developing and implementing quantum programs across broad user communities. Upon examining the current landscape of quantum software for mathematical optimization, we observe that the majority of the software dedicated to quantum optimization focuses on addressing combinatorial and discrete optimization problems, with limited options available for continuous optimization.

Generally, a combinatorial optimization problem can be reformulated as a Quadratically Unconstrained Binary Optimization (QUBO) problem, the solution of which is believed to be a promising application of quantum computing~\citep{quintero2022qubo}.
There is a rich collection of libraries for quantum and quantum-inspired optimization that can be employed to generate QUBO reformulations, including \texttt{QUBO.jl}~\citep{xavier2023qubo}, \texttt{Amplify}~\citep{matsuda2020research}, \texttt{PyQUBO}~\citep{zaman2021pyqubo}, \texttt{qubovert}~\citep{iosue2022qubovert}. 
These QUBO problems can be tackled by several methods, such as quantum annealing, Quantum Approximate Optimization Algorithms (QAOA)~\citep{farhi2014quantum}, and other hybrid approaches~\citep{yamamoto2017coherent}. 
D-Wave's \texttt{Ocean SDK}~\citep{dwave2023ocean} enables users to interface with their direct QPU (i.e., quantum annealer) and hybrid solvers and retrieve results.
QuEra's \texttt{Bloqade.jl}~\citep{quera2024bloqade} is a high-level language for configuring programmable Rydberg atom arrays, which can be used to implement annealing-type quantum algorithms and discrete optimization problems like QUBO~\citep{nguyen2023quantum}.
Los Alamos Advanced Network Science Initiative has also released a package named \texttt{QuantumAnnealing.jl}~\citep{morrell2022quantum} for simulation and execution of quantum annealing. 
Besides, several software packages have been published for programming quantum circuits, including IBM's \texttt{Qiskit}~\citep{aleksandrowicz2019qiskit}, Google's \texttt{Cirq}~\citep{developers2019cirq}, Amazon's \texttt{Braket SDK}~\citep{aws2024braket}, Microsoft's \texttt{Q\#}~\citep{singhal2022q}, and Xanadu's \texttt{PennyLane}~\citep{bergholm2018pennylane}. These tools can be used to deploy QAOA on gate-based quantum computers.

While there have been a few proposals for solving continuous optimization problems using quantum or hybrid computing devices such as photonic quantum computers~\citep{verdon2019quantum} and coherent continuous variable machines~\citep{khosravi2022non}, we are not aware of a software library customized for nonlinear continuous optimization problems. In practice, some nonlinear optimization problems, such as quadratic programming, may be reformulated as QUBO problems and handled by the aforementioned software tools. However, it remains unclear whether this approach could lead to robust quantum advantages.

\section{Comparison with Existing Tools}\label{sec:comparison}

As we discussed in \sec{workflow}, \qhdopt\ first obtains some low-resolution solutions by executing the QHD algorithm for a nonlinear optimization problem through Hamiltonian embedding.
Next, the software employs a classical local search strategy for fast post-processing of the raw quantum results. It is of interest to understand to what extent the quantum component (i.e., the noisy implementation of QHD) improves the overall performance of \qhdopt. To this end, we have designed a benchmark test to evaluate the performance of \qhdopt~for nonlinear and nonconvex optimization problems.

\subsection{Test problems}
We demonstrate the performance of \qhdopt~using fifteen randomly generated nonlinear optimization instances, all with unit box constraints.
Problem instances 1 - 5 are nonlinear programming (NLP) problems involving two or three continuous variables, as detailed in \tab{test-info}.
Problem instances 6 - 10 are quadratic programming (QP) problems drawn from the benchmark devised in \citep{leng2023quantum}.
Problems instances 11 - 15 are nonlinear programming (NLP) problems involving exponential functions, as specified in the following expression:
\begin{equation}
    f(x) = \frac{1}{2}\sum^{N}_{i=1} \sum^{N}_{j=1}Q_{i,j}e^{x_i}e^{x_j} + \sum^N_{i=1} b_i e^{-x_i}.
\end{equation}
The last ten test instances (6 - 15) are intermediate-scale problems with 50 continuous variables, ranging from $0$ to $1$. To ensure the successful mapping of these test problems to quantum computers with limited connectivity, these test problems are generated in a way such that their Hessians are sparse matrices.\footnote{Detailed expressions of these test instances are provided in the software repository, see the ``examples'' folder.}

\begin{table}[!ht]
    \centering
    \begin{tabular}{|p{3cm}|p{14cm}|}
     \hline
     \textbf{Test index} & \textbf{Problem description}\\
     \hline
     1 & $f(x,y) = -4x^2 + 3xy - 2y^2 + 3x - y$\\
     \hline
     2 & $f(x,y) = -2\left(x-\frac{1}{3}\right)^2 + y^2 -\frac{1}{3}y\log(3x + \frac{1}{2}) + 5\left(x^2 - y^2 - x - \frac{1}{2}\right)^2$\\
     \hline
     3 & $f(x,y)=y^{3/2} - e^{4x}\left(y-\frac{3}{4}\right)$ \\
     \hline
     4 & $f(x,y,z)  = (2y-1)^2\left(z-\frac{2}{5}\right) - (2x-1)z + y\left(2x-\frac{3}{2}\right)^2$\\
     \hline
     5 & $f(x,y,z)  = 2e^{-x} * (2z-1)^2 - 3\left(2y-\frac{7}{10}\right)^2 e^{-z} + \log(x + 1) \left(y-\frac{4}{5}\right)$\\
     \hline
    \end{tabular}
    \caption{Problem instances 1 - 5 for nonlinear programming. All the test instances are nonconvex problems with unit box constraints $[0,1]^n$.}
    \label{tab:test-info}
\end{table}

These instances were generated in a largely random manner, and each possesses multiple local solutions, making them fairly challenging for classical optimization software. In our experiment, we observed that local solvers, such as Ipopt, cannot find globally optimal (or even approximately optimal) solutions unless a large number of random initial guesses are tried. BARON, a highly optimized commercial solver for global optimization, can find globally optimal solutions to sparse quadratic programming problems in under 1 second, but it takes several minutes to certify global optimality for nonlinear programming problems that involve exponential-type objectives.

\subsection{Experiment setup and results}
In this subsection, we discuss the basic setup of the experiment and the numerical results. We test \qhdopt\ with two different post-processing optimizers (i.e., Ipopt and Scipy-TNC) using the randomly generated nonlinear programming instances discussed in the previous section. As a comparison, we also run the two classical optimizers on the same test instances using uniformly random initialization.
These two classical optimizers are assessed as baselines to illustrate the quantum advantage introduced by the D-Wave-implemented Quantum Hamiltonian Descent (QHD).
The classical components in both experiments, including the decoding of D-Wave samples and classical refinement, were executed on a 2022 MacBook Pro laptop with an Apple M2 chip.
Our findings assert that QHD, when implemented with D-Wave, brings a significant advantage compared to the standalone use of classical optimizers.

\subsubsection{Experiment setup for \qhdopt}
We evaluate \qhdopt\ on this benchmark using the D-Wave Advantage\_system6.3 as the quantum backend. For a fair comparison, we use the unary embedding scheme for all instances, including quadratic and non-quadratic problems.
The anneal time is set to be the default value, i.e., 20 microseconds. 
The total quantum runtime per shot (see the ``QPU'' columns in \tab{benchmark}) is calculated as the arithmetic mean of the ``qpu\_access\_time'' reported by D-Wave, which includes the programming, state preparation, annealing, and decoding. Note that we do not include the transmission time and the task queuing time in our report.
We test \qhdopt\ with two post-processing optimizers (i.e., Ipopt and Scipy-TNC), and the classical post-processing (more precisely, classical refinement) time is reported in the ``Classical Refine'' columns in \tab{benchmark}. The standard deviation of the classical refinement time is also reported in the parenthesis. Except for the initial guesses, both solvers use the default parameters as provided with their Python API.

\subsubsection{Baseline using classical optimizers}
As a comparison, we also test three classical optimizers for the same set of problems: Ipopt, Scipy-TNC, and BARON. The first two optimizers are initialized with 1000 uniformly random guesses in the unit box $[0,1]^d$ (where $d$ is the problem dimension), and the runtime data for the 1000 runs have been collected. For a fair comparison, we use the same random seeds for both methods. BARON is executed to generate global solutions with a 2-minute timeout. Except for the initial guesses, all solvers use the default parameters as provided with their Python API. \tab{control-group} shows the runtime of the three classical solvers: for Ipopt and TNC, the arithmetic mean (and standard deviation) of the runtime is reported; for BARON, the total runtime is reported. 

Note that for the last five test instances, BARON failed to certify the global optimality of the obtained solutions within the 2-minute timeout window. In \tab{baron-sol-quality}, we further investigate BARON's solution quality. Our results suggest that, while BARON can find solutions as good as those from the other tested solvers in a comparable timescale, a much longer time is required to prove the global optimality of the obtained solutions. Therefore, we regard the solution returned by BARON as the global minimum.

\subsubsection{Performance metric}
In the experiments, we use \textit{time-to-solution} (TTS) as the key metric to evaluate the performance of various optimization methods. 
TTS is defined as the total runtime required by a method to achieve at least $0.99$ success probability. It can be calculated using the following formula,
$$\text{TTS} = t_0 \times \left\lceil\frac{\ln(1-0.99)}{\ln(1-p_s)}\right\rceil,$$
where $t_0$ is the (average) runtime per shot, and $p_s$ is the success probability.
For all the 15 test instances, the time-to-solution data of four methods (QHD+Ipopt, QHD+TNC, Ipopt, and TNC) are presented in \tab{tts-summary}.
For the experiments involving QHD (i.e., the results in \tab{benchmark}), $t_0$ is calculated as the sum of average QPU time and average classical refinement time; for the experiments that only involve classical optimizers (i.e., the results in \tab{control-group}), $t_0$ is equivalent to the (average) classical runtime. 
The success probability $p_s$ is estimated by the fraction of ``successful'' events in the 1000 samples/trials. Here, a result $x'$ is considered successful if the optimality gap $f(x') -f(x^*)$ is less than $0.001$, where $x^*$ is the solution obtained by BARON.

\begin{table}[!ht]
    \centering
    \begin{adjustbox}{width = 1\linewidth}
    \begin{tabular}{|P{1.5cm}|P{2.5cm}|P{2.5cm}|P{2cm}|P{2cm}||P{1.5cm}|P{2.5cm}|P{2.5cm}|P{2cm}|P{2cm}|}
     \hline
     \textbf{Test index} & \textbf{QHD+Ipopt} & \textbf{QHD+TNC} & \textbf{Ipopt} & \textbf{TNC} & \textbf{Test index} & \textbf{QHD+Ipopt} & \textbf{QHD+TNC} & \textbf{Ipopt} & \textbf{TNC}\\
     \hline
     1 & 2.05e-1 & \underline{3.68e-2} & 7.19e+1 & 1.57e+0 & 9 & 1.82e-1 & \underline{4.44e-2} & 3.17e+1 & 2.68e+1\\
     \hline
     2 & 5.69e-1 & \underline{1.11e-1} & 7.56e+1 & 3.73e+0 & 10 & 8.94e-1 & \underline{2.06e-1} & 7.65e+0 & 5.50e+0\\
     \hline
     3 & 5.50e-1 & \underline{3.52e-2} & 8.51e+1 & 3.83e+0 & 11 & 1.21e+0 & \underline{1.17e-2} & 1.98e-1 & 3.65e-2\\
     \hline
     4 & 4.60e+0 & \underline{2.95e-1} & 3.20e+1 & 2.10e+0 & 12 & 7.57e+0 & \underline{3.32e-2} & 4.15e+0 & 2.22e-1\\
     \hline
     5 & 5.14e-1 & \underline{6.73e-2} & 6.67e+1 & 3.03e+0 & 13 & 3.78e+2 & \underline{2.80e-1} & 3.52e+1 & 1.38e-1\\
     \hline
     6 & 3.74e-1 & \underline{6.51e-2} & 9.41e-1 & 5.12e+0 & 14 & 6.06e-1 & \underline{3.56e-2} & 7.82e+0 & 3.92e-1\\
     \hline
     7 & 9.42e-1 & \underline{1.49e-1} & 3.12e+1 & 1.15e+1 & 15 & 1.36e+0 & \underline{3.26e-2} & 9.34e+0 & 1.36e-1\\
     \hline
     8 & 3.96e-2 & \underline{5.14e-3} & 4.30e-1 & 9.57e-1 & & & & & \\
     \hline
    \end{tabular}
    \end{adjustbox}
    \caption{Time-to-solution data of all four methods (unit: second). The lowest TTS is underlined per instance.}
    \label{tab:tts-summary}
\end{table}

\begin{table}[!ht]
    \tiny
    \centering
    \begin{adjustbox}{width = 1\linewidth}
    \begin{tabular}{|c|p{3.1em}|p{7.5em}|p{3.1em}|>{\columncolor[gray]{0.9}}p{3.1em}||p{3.1em}|p{7.5em}|p{3.1em}|>{\columncolor[gray]{0.9}}p{3.1em}|}
    \hline
    \multirow{2}{1.7em}{Test\\Index} & \multicolumn{4}{c||}{\textbf{QHD+Ipopt}} & \multicolumn{4}{c|}{\textbf{QHD+TNC}}\\
    \cline{2-9}
         & QPU & Classical Refine & SP & TTS & QPU & Classical Refine & SP & TTS\\
        \hline
        1 & 1.15e-3 & 2.03e-1 (5.87e-4) & 9.96e-1 & 2.05e-1 & 1.15e-3 & 3.19e-2 (1.41e-5) & 9.84e-1 & \underline{3.68e-2}\\
        \hline
        2 & 9.97e-4 & 3.16e-1 (8.08e-3) & 9.23e-1 & 5.69e-1 & 9.97e-4 & 5.24e-2 (1.16e-4) & 9.12e-1 & \underline{1.11e-1}\\
        \hline
        3 & 1.08e-3 & 3.35e-1 (2.88e-3) & 9.40e-1 & 5.50e-1 & 1.08e-3 & 2.96e-2 (1.17e-4) & 9.82e-1 & \underline{3.52e-2}\\
        \hline
        4 & 1.25e-3 & 1.6e+0 (5.48e-3) & 7.98e-1 & 4.60e+0 & 1.25e-3 & 1.28e-1 (2.68e-4) & 8.67e-1 & \underline{2.95e-1}\\
        \hline
        5 & 1.23e-3 & 3.64e-1 (1.08e-3) & 9.62e-1 & 5.14e-1 & 1.23e-3 & 5.75e-2 (3.34e-5) & 9.82e-1 & \underline{6.73e-2}\\
        \hline
        6 & 1.69e-3 & 2.03e-2 (5.96e-3) & 2.39e-1 & 3.74e-1 & 1.69e-3 & 1.57e-3 (4.85e-4) & 2.08e-1 & \underline{6.51e-2}\\
        \hline
        7 & 2.04e-3 & 2.34e-2 (1.64e-3) & 1.20e-1 & 9.42e-1 & 2.04e-3 & 1.99e-3 (4.92e-4) & 1.20e-1 & \underline{1.49e-1}\\
        \hline
        8 & 1.49e-3 & 1.83e-2 (1.93e-3) & 9.23e-1 & 3.96e-2 & 1.49e-3 & 1.08e-3 (2.81e-4) & 9.78e-1 & \underline{5.14e-3}\\
        \hline
        9 & 2.01e-3 & 1.81e-2 (3.90e-3) & 4.36e-1 & 1.82e-1 & 2.01e-3 & 2.35e-3 (3.48e-4) & 3.92e-1 & \underline{4.44e-2}\\
        \hline
        10 & 2.11e-3 & 2.02e-2 (5.62e-3) & 1.09e-1 & 8.94e-1 & 2.11e-3 & 2.91e-3 (5.21e-4) & 1.06e-1 & \underline{2.06e-1}\\
        \hline
        11 & 2.04e-3 & 9.91e-2 (2.24e-2) & 3.42e-1 & 1.21e+0 & 2.04e-3 & 9.69e-3 (6.67e-4) & 9.97e-1 & \underline{1.17e-2}\\
        \hline
        12 & 2.14e-3 & 9.25e-2 (1.38e-2) & 4.45e-1 & 7.57e+0 & 2.14e-3 & 8.94e-3 (1.19e-3) & 8.67e-1 & \underline{3.32e-2}\\
        \hline
        13 & 2.09e-3 & 1.62e-1 (2.08e-2) & 2e-3 & 3.78e+2 & 2.09e-3 & 1.19e-2 (8.99e-4) & 2.11e-1 & \underline{2.80e-1}\\
        \hline
        14 & 1.94e-3 & 5.89e-2 (1.33e-2) & 3.83e-1 & 6.06e-1 & 1.94e-3 & 6.94e-3 (7.84e-4) & 7.46e-1 & \underline{3.56e-2}\\
        \hline
        15 & 2.13e-3 & 8.26e-2 (1.13e-2) & 2.53e-1 & 1.36e+0 & 2.13e-3 & 8.75e-3 (6.67e-4) & 8.88e-1 & \underline{3.26e-2}\\
        \hline
    \end{tabular}
    \end{adjustbox}
    \caption{Performance of \qhdopt\ on 15 randomly generated nonlinear programming test instances. ``SP'' represents success probability. The units of quantum runtime (i.e., ``QPU''), classical post-processing time (i.e., ``Classical Refine''), and time-to-solution (i.e., ``TTS'') are second. The standard deviation of classical post-processing time is shown in the parenthesis. The lowest TTS in a row is underlined.}
    \label{tab:benchmark}
\end{table}

\begin{table}[!ht]
    \tiny
    \centering
    \begin{adjustbox}{width = 1\linewidth}
    \begin{tabular}{|c|p{7.6em}|p{4em}|>{\columncolor[gray]{0.9}}p{4em}||p{7.6em}|p{4em}|>{\columncolor[gray]{0.9}}p{4em}||p{4em}|}
    \hline
    \multirow{2}{1.7em}{Test\\Index} & \multicolumn{3}{c||}{\textbf{Ipopt}} & \multicolumn{3}{c||}{\textbf{TNC}} & \multicolumn{1}{c|}{\textbf{BARON}}\\
    \cline{2-8}
         & Avg. Runtime & SP & TTS & Avg. Runtime & SP & TTS & Runtime\\
        \hline
        1 & 1.33e+1 (2.22e-3) & 5.72e-1 & 7.19e+1 & 2.82e-1 (1.29e-5) & 5.64e-1 & \underline{1.57e+0} & 1.00e-2\\ 
        \hline
        2 & 1.29e+1 (4.87e-3) & 5.44e-1 & 7.56e+1 & 5.86e-1 (2.72e-4) & 5.15e-1 & \underline{3.73e+0} & 1.00e-2\\
        \hline
        3 & 2.09e+1 (1.61e-2) & 6.78e-1 & 8.51e+1 & 6.85e-1 (1.08e-4) & 5.61e-1 & \underline{3.83e+0} & 1.00e-2\\
        \hline
        4 & 9.74e+0 (4.71e-3) & 7.54e-1 & 3.20e+1 & 5.31e-1 (2.68e-4) & 6.87e-1 & \underline{2.10e+0} & 1.00e-2\\
        \hline
        5 & 1.43e+1 (2.78e-3) & 6.27e-1 & 6.67e+1 & 6.42e-1 (3.30e-5) & 6.23e-1 & \underline{3.03e+0} & 1.00e-2\\
        \hline
        6 & 3.77e-2 (1.39e-2) & 1.70e-1 & \underline{9.41e-1} & 6.69e-3 (1.47e-3) & 6.00e-3 &5.12e+0 & 1.90e-1\\
        \hline
        7 & 5.44e-2 (1.50e-2) & 8.00e-3 & 3.12e+1 & 5.00e-3 (7.01e-4) & 2.00e-3 & \underline{1.15e+1} & 4.00e-2\\
        \hline 
        8 & 4.78e-2 (1.39e-2) & 4.05e-1 & \underline{4.30e-1} & 5.04e-3 (1.59e-3) & 2.40e-2 & 9.57e-1 & 3.00e-2\\
        \hline 
        9 & 4.83e-2 (1.22e-2) & 7.00e-3 & 3.17e+1 & 5.83e-3 (8.76e-4) & 1.00e-3 & \underline{2.68e+1} & 3.00e-2\\
        \hline
        10 & 5.38e-2 (2.14e-2) & 3.20e-2 & 7.65e+0 &7.18e-3 (1.03e-3) & 6.00e-3 & \underline{5.50e+0} & 5.00e-2\\
        \hline 
        11 & 9.88e-2 (3.23e-2) & 9.28e-1 & 1.98e-1 & 9.13e-3 (1.01e-3) & 7.00e-1 &\underline{3.65e-2} & 1.20e+2\\
        \hline
        12 & 1.15e-1 (2.70e-2) & 1.21e-1 & 4.15e+0 & 8.88e-3 (1.27e-3) & 1.71e-1 &\underline{2.22e-1} & 1.20e+2\\
        \hline 
        13 & 9.99e-2 (2.65e-2) & 1.3e-2 &3.52e+1 &9.19e-3 (1.08e-3) &2.77e-1 &\underline{1.38e-1} & 1.20e+2\\
        \hline 
        14 & 1.10e-1 (4.73e-2) & 6.30e-2 & 7.82e+0 & 9.56e-3 (1.15e-3) & 1.08e-1 & \underline{3.92e-1} & 1.20e+2\\
        \hline 
        15 & 9.07e-2 (2.81e-2) & 4.40e-2 &9.34e+0 &9.06e-3 (1.02e-3) & 2.74e-1 & \underline{1.36e-1} & 1.20e+2\\
        \hline
    \end{tabular}
    \end{adjustbox}
    \caption{Performance of classical optimizers on the same 15 randomly generated test instances as a baseline. ``SP'' represents success probability. The units of classical optimizer runtime time (i.e., ``Avg. Runtime'' and ``Runtime'') and time-to-solution (i.e., ``TTS'') are seconds. The standard deviation of runtime is shown in the parenthesis. The lowest TTS in a row is underlined.}
    \label{tab:control-group}
\end{table}

\begin{table}[!ht]
    \centering
    \begin{adjustbox}{width = 0.97\linewidth}
    \begin{tabular}{|P{1.5cm}|P{2.5cm}|P{3cm}|P{3cm}||P{1.5cm}|P{2.5cm}|P{3cm}|P{3cm}|}
     \hline
     \textbf{Test index} & \textbf{Best found obj.} & \textbf{BARON result 1} & \textbf{BARON result 2} & \textbf{Test index} & \textbf{Best found obj.} & \textbf{BARON result 1} & \textbf{BARON result 2}\\
     \hline
     1 & -3 & -3 & -3 & 9 & -2.269 & -2.269 & -2.269\\
     \hline
     2 & 0.354 & 0.354 & 0.354 & 10 & -2.305 & -2.305 & -2.305\\
     \hline
     3 & -12.650 & -12.650 & -12.650 & 11 & -31.256 & -31.256 & -31.256\\
     \hline
     4 & -0.882 & -0.882 & -0.882 & 12 & -66.618 & -66.618 & -66.618\\
     \hline
     5 & -4.196 & -4.196 & -4.196 & 13 & -56.762 & -56.762 & -56.762\\
     \hline
     6 & -1.188 & -1.188 & -1.188 & 14 & -25.357 & -25.357 & - 25.357\\
     \hline
     7 & -2.110 & -2.110 & -2.110 & 15 & -59.342 & -59.188 & -59.342\\
     \hline
     8 & -1.809 & -1.809 & -1.809 & & & & \\
     \hline
    \end{tabular}
    \end{adjustbox}
    \caption{Performance of BARON. BARON result 1 (or 2) indicates BARON's best-found objective function value given a runtime no longer than TNC's (or Ipopt's) time-to-solution as reported in \tab{control-group}. Except for instance 15, BARON always finds the optimal solution in the given time windows.}
    \label{tab:baron-sol-quality}
\end{table}

\subsection{Interpretation of the experiment results}

Based on the time-to-solution data as reported in \tab{tts-summary}, we observed that \qhdopt\ (QHD + a classical optimizer) always outperforms the standalone use of a classical optimizer for the 15 randomly generated test instances. As for the two local optimizers, we find that Scipy-TNC works better than Ipopt as a post-processing subroutine. While the two optimizers usually return refined samples with comparable success probability, Scipy-TNC always shows a lower TTS due to a notably faster runtime. This is potentially because TNC is a quasi-Newton method that does not need to solve the full Newton linear system in the iterations.

\begin{figure*}
    \centering
    \includegraphics[width=0.95\linewidth, trim = 3cm 0cm 3cm 0cm, clip]{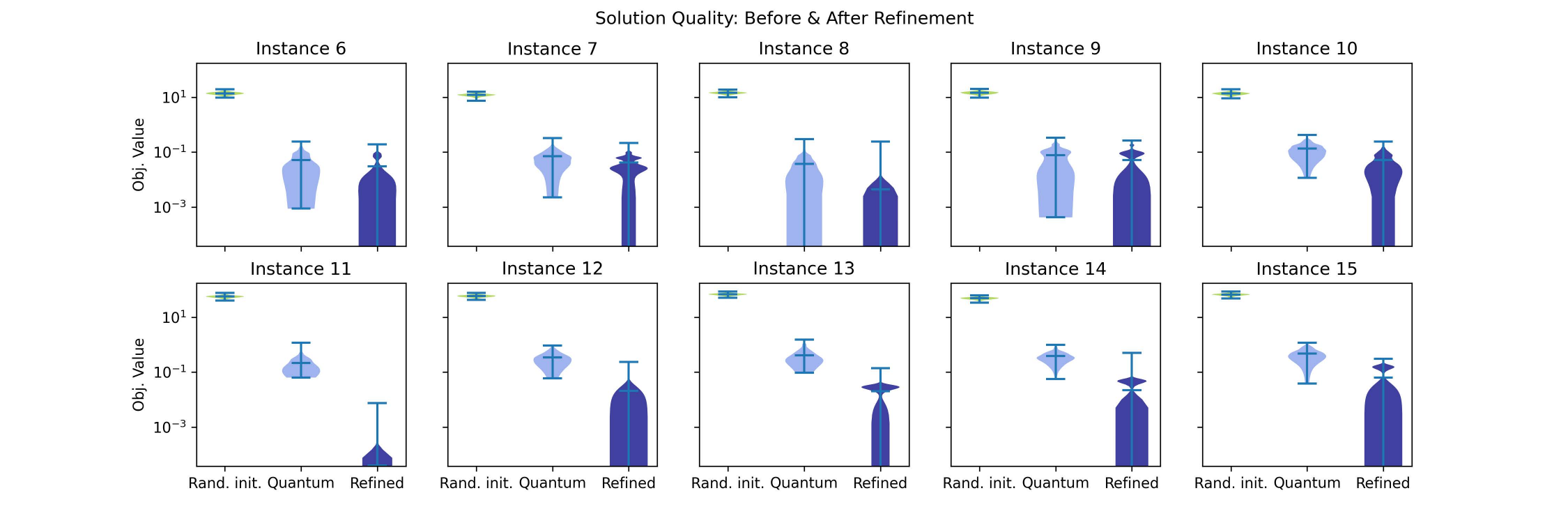}
    \caption{\small{Comparison of solution quality using randomly generated initial guesses, quantum-generated samples, and classically refined solutions.}}
    \label{fig:solution-quality}
\end{figure*}

Another interesting finding is that the classical refinement times of quantum-generated samples (see ``Classical Refine'' in \tab{benchmark}) in \qhdopt\ are significantly shorter than the average runtime of the direct use of local optimizers (see ``Avg. Runtime'' in \tab{control-group}). For example, in test instance 1, the average post-processing time (using Ipopt) for a quantum-generated initial guess is 0.2 s, while the average runtime of Ipopt given uniformly random guesses is 13 s.
To further investigate this phenomenon, we plot the distribution of objective function values corresponding to three different sample groups, including (1) randomly generated initial guesses, (2) quantum (D-Wave) generated samples, and (3) TNC refined samples (using quantum-generated samples as initial guesses), as shown in~\fig{solution-quality}.\footnote{In~\fig{solution-quality}, we only plot objective function values for high-dimensional problems, i.e., test instances 6 - 15.} 
While the quantum-generated solutions are limited by low precision, it is observed that they are still qualitatively better than random initial guesses. In the subsequent post-processing, \qhdopt\ performs a local search subroutine to refine solution quality by improving numerical accuracy. In other words, the quantum sampler in \qhdopt\ can be regarded as a fast and efficient warm-start that devises initial guesses of better quality.

\section{Conclusion and Future Work}
\qhdopt\ is the first open-source software leveraging quantum devices for nonconvex nonlinear optimization problems, providing an accessible interface for domain experts without quantum computing knowledge. Exploiting the idea of Hamiltonian-oriented programming, it efficiently uses quantum devices by implementing the Quantum Hamiltonian Descent algorithm with the \simuq\ framework. We demonstrated \qhdopt's effectiveness through examples and benchmarks, showing its advantage over classical solvers, especially in large, complex instances. However, the current limitations of quantum device programmability and scalability constrain our benchmarks' scale. While QHD shows promise in solving complex optimization problems, further empirical studies are needed for real-world performance evaluation.

There are several avenues for future development of \qhdopt. First, it is desired to broaden the problem class that can be handled by \qhdopt.
Currently, due to hardware limitations, \qhdopt\ supports only the optimization of box-constrained nonlinear problems defined as a sum of univariate and bivariate functions. We anticipate that, shortly, \qhdopt\ can be extended to address more complicated objective functions as quantum technology and quantum algorithm design continue to co-evolve.
Second, while local search algorithms work well to improve the precision of quantum-generated samples, to obtain a global optimality guarantee, it might be promising to replace the refinement/post-processing subroutine in \qhdopt\ with global optimizers (for example, those based on branch-and-bound).
Third, with further progress in quantum engineering, \qhdopt\ is expected to support more quantum devices from different platforms, including commercial or laboratory devices, which is essential to understanding the advantage of \qhdopt\ given different combinations of embedding schemes and quantum devices. Last but not least,
\qhdopt\ can be expanded into a plugin for various domain-specific tools, including those in engineering, management, finance, and economics. Adaptions to specific domains are invaluable for users to better utilize quantum devices for their domain problems.  
Our overarching goal is to establish a user-friendly tool, empowering individuals and organizations to harness the power of quantum devices to solve challenging problems in the real world.

\section*{Acknowledgment}
This work was partially funded by the U.S. Department of Energy, Office of Science, Office of Advanced Scientific
Computing Research, Accelerated Research in Quantum Computing under Award Number DE-SC0020273, the Air
Force Office of Scientific Research under Grant No. FA95502110051, the U.S. National Science Foundation grant CCF-1816695, CCF-1942837 (CAREER), ECCS-2045978, a Sloan research fellowship, the Simons Quantum Postdoctoral Fellowship, and a Simons Investigator award through Grant No. 825053. J.L. and Y.P. are also supported by an open-source quantum software grant from the Unitary Fund.

%%%%%%%%%%%%%%%%%%%%%%%%%%%%%%%%%%%%%%%%%%%%%%%%%%%%%%%%%%%%%%%

\bibliographystyle{plainnat}
\bibliography{refs}

\end{document}